# Spatially resolved ultrafast magnetic dynamics launched at a complex-oxide hetero-interface

M. Först[1*], A.D. Caviglia[2*], R. Scherwitzl[3], R. Mankowsky[1], P. Zubko[3], V. Khanna[1,4,5], H. Bromberger[1], S.B. Wilkins[6], Y.-D. Chuang[7], W.S. Lee[8], W.F. Schlotter[9], J.J. Turner[9], G.L. Dakovski[9], M.P. Minitti[9], J. Robinson[9], S.R. Clark[10,4], D. Jaksch[4,10], J.-M. Triscone[3], J.P. Hill[6], S.S. Dhesi[5], and A. Cavalleri[1,4]

[1]Max-Planck Institute for the Structure and Dynamics of Matter, Hamburg, Germany

[2]Kavli Institute of Nanoscience, Delft University of Technology, The Netherlands

[3]Département de Physique de la Matière Condensée, University of Geneva, Switzerland

[4]Department of Physics, Clarendon Laboratory, University of Oxford, UK

[5]Diamond Light Source, Chilton, Didcot, United Kingdom

[6]Condensed Matter Physics and Materials Science Department, Brookhaven National Laboratory, Upton, NY

[7]Advanced Light Source, Lawrence Berkeley Laboratory, Berkeley, CA

[8] The Stanford Institute for Materials and Energy Sciences (SIMES), Stanford Linear Accelerator Center (SLAC) National Accelerator Laboratory and Stanford University, Menlo Park, CA

[9] Linac Coherent Light Source, Stanford Linear Accelerator Center (SLAC) National Accelerator Laboratory, Menlo Park, CA

[10] Centre for Quantum Technologies, National University of Singapore, Singapore

[*]These authors contributed equally to this work.

**Static strain in complex oxide heterostructures [1,2] has been extensively used to engineer electronic and magnetic properties at equilibrium [3]. In the same spirit, deformations of the crystal lattice with light may be used to achieve functional control across hetero-interfaces dynamically [4]. Here, by exciting large amplitude infrared-active vibrations in a $LaAlO_3$ substrate we induce magnetic order melting in a $NdNiO_3$ film across a hetero-interface. Femtosecond Resonant Soft X-ray Diffraction is used to determine the spatial and temporal evolution of the magnetic disordering. We observe a magnetic melt front that grows from the substrate interface into the film, at a speed that suggests electronically driven propagation. Light control and ultrafast phase front propagation at hetero-interfaces may lead to new opportunities in optomagnetism, for example by driving domain wall motion to transport information across suitably designed devices.**

In transition metal oxides, rearrangements in electronic and magnetic properties can be triggered by the application of magnetic [5] and electric fields [6], or pressure [7,8]. Switching has also been demonstrated in these materials using femtosecond optical excitation, at near-visible [9,10,11,12,13,14,15], mid-infrared [16,17,18,19,20], or THz [21,22,23] wavelengths.

Recently, selective excitation of lattice modes in the mid-infrared has been applied to complex oxide heterostructures, where the functional material can be separated from the optically excited region. By directly driving the infrared-active modes leading to structural distortions in the $LaAlO_3$ substrate of a $LaAlO_3/NdNiO_3$ heterostructure, an insulator-metal transition was triggered across the interface in the nickelate film [4].

Here, we apply time-resolved Resonant Soft X-ray Diffraction (RSXD) to measure the concomitant magnetic response in the $NdNiO_3$ film with nanometer spatial and femtosecond temporal resolution. We find evidence of inhomogeneous magnetic melting dynamics, with a melt front that propagates from the interface into the $NdNiO_3$ film at speeds of order of, or even in excess of, the speed of sound. An insulator-metal transition initiated at the hetero-interface and propagating inward, possibly locked to a wave of local octahedral distortions, are considered to explain these observations. Theoretical estimations indicate that itinerant charge carriers scramble the magnetic order efficiently as they move into the film.

Below about 200 K, metallic paramagnetic $NdNiO_3$ undergoes a transition into a low-temperature antiferromagnetic insulating state [24,25]. This electronic and magnetic phase transition is concomitant with a structural transformation from an orthorhombic (Pbnm) to a monoclinic (P21/n) crystal structure. Further, the transition temperature depends on epitaxial strain [26], demonstrating sensitivity to lattice distortions.

We study a compressively strained film of 100 NdNiO$_3$ unit cells deposited on a (111) LaAlO$_3$ substrate by off-axis RF magnetron sputtering [27]. The low-temperature antiferromagnetic ordering on the Ni and Nd sublattices, shown schematically in Fig. 1(a), can be observed by RSXD at the pseudo-cubic (1/4 1/4 1/4) wave vector [28,29]. Figure 1(b) shows the static photon-energy dependent magnetic diffraction at this peak, measured at the Ni L$_{2,3}$ edges (data taken at I06 beamline of the Diamond Light Source synchrotron; see Ref. 30 for details).

Femtosecond RSXD experiments were carried out at the SXR beamline of the Linac Coherent Light Source free electron laser (FEL) [31]. Excitation pulses of 200 fs duration were tuned to 15 μm wavelength (82 meV photon energy), in resonance with the highest-frequency LaAlO$_3$ substrate phonon, which is well separated in energy from equivalent phonons of the NdNiO$_3$ film (see Figure 1(c)). We focused these mid-infrared pump pulses onto the sample with a fluence of 4 mJ/cm$^2$. The sample was mounted on an in-vacuum diffractometer [32] and cooled to 40 K, i.e. into the antiferromagnetic insulating state. The FEL, which operated at 120 Hz repetition rate, was tuned to the 852 eV Ni L$_3$ edge. The bandwidth of the X-ray pulses was reduced to below 1 eV by a grating monochromator. Diffracted X-rays were detected as function of the time delay relative to the mid-infrared excitation pulses. An avalanche photodiode enabled pulse-to-pulse normalization of the diffracted to the incident light intensity.

Figure 1(d) shows the vibrationally induced intensity changes of the (1/4 1/4 1/4) diffraction peak, probing the antiferromagnetic order dynamics in the NdNiO$_3$ film. The peak intensity dropped by about 80 % within 1.6 ps and recovered on the few tens of picosecond time scale. In contrast to the case of direct interband excitation, for which the antiferromagnetic peak dropped promptly [30], the 1.6 ps time constant for the initial reduction was significantly longer than the 250 fs time resolution of the experiment, where the latter is limited by the jitter between the FEL and the optical laser. We compare these dynamics to the time needed for the film to become metallic, as measured by the transient reflectivity in the 1–5 THz range induced by the same mid-infrared excitation (green dots in Fig. 1(d)). These two similar timescales, which reflect only average changes over the whole film, suggest an intimate connection between the insulator-to-metal transition and the melting of magnetic order.

In Figure 2, we plot the transient $\theta$-$2\theta$ scans for the (1/4 1/4 1/4) diffraction peak, sensitive to the out-of-plane antiferromagnetic ordering. In equilibrium, i.e. at negative time delay, a narrow diffraction peak and Laue oscillations are observed, attesting to the presence of magnetic order across the entire 30-nm film height, with sharp magnetic boundaries.

Figure 2 further shows that a significant peak broadening and a suppression of the Laue oscillations accompany the strong photo-induced reduction in peak intensity. The broadening of the diffraction peak implies that the excitation melts the magnetic order only over a fraction of the film along the sample growth direction. Secondly, the suppression of the Laue oscillations indicates that the boundary between the ordered and disordered regions of the film is not sharp.

We also find that throughout these dynamics the in-plane correlation length, as measured by transverse rocking curves ($\theta$ scans), remains unchanged (see Supplementary

Information). Hence, the dynamics discussed here are one dimensional, evolving along the sample growth direction.

The spatial distribution of the magnetic order at a time delay $\tau$ was analyzed quantitatively with the following expression for kinematic diffraction

$$I_\tau(q) \propto \left| \int_0^D F(z,\tau) e^{-iqz} dz \right|^2. \tag{1}$$

Here, the magnetic profile is represented by the space- and time-dependent structure factor $F(z,\tau)$, where $q = 4\pi \sin\theta/\lambda$ is the magnitude of the scattering wave vector (with $\theta$ the diffraction angle and $\lambda$ the x-ray wavelength), $z$ the distance into the film, and $D$ the film thickness.

Previous work has shown that these complex dynamics are initiated at the buried interface [4]. Also, the transient $\theta$-$2\theta$ scans show that the fraction of material, which is melted, increases with time and that the order-disorder interface becomes progressively smeared in time. Hence we chose a functional form $F(z,\tau)$ that describes a soliton-like demagnetization front, propagating from the hetero-interface into the thin film with a smeared and adjustable phase front:

$$F(z,\tau) = F_0 \cdot \left( \frac{1}{2} + \frac{1}{2}\text{erf}(\frac{z-z_f(\tau)}{d_f(\tau)}) \right). \tag{2}$$

In this case, $F_0$ is the equilibrium structure factor, and $z_f$ and $d_f$ are the time-dependent position and width of that phase front separating the unperturbed antiferromagnetic order from the disordered region of the film. Numerical fits of Eqs. (1) and (2) to the diffraction peak at selected time delays are shown as red solid lines in the lower panel of Fig. 2 and give excellent agreement with the data.

Figure 3(b) shows the corresponding early time scale evolution of the space-dependent magnetic order parameter, represented as $|F(z,\tau)|^2$. At negative time delays, the NdNiO$_3$ film is homogenously ordered, with a sharp boundary at the hetero-interface. The mid-infrared excitation induces heterogeneous melting, with a demagnetizing phase front that propagates halfway into the film – before coming to a stop at ~2 ps. This phase front leaves a magnetically disordered region behind, with a boundary between the two regions of about 10 nm width. The time-dependent position of the phase front $z_f$ is plotted in Figure 3(c), which suggests that the magnetic melt front propagates at a speed of order of, or even faster than the longitudinal sound velocity measured in an NdNiO$_3$ film [33].

The re-magnetization dynamics are shown in Figure 4. About 10 ps after the excitation, the demagnetized part of the film starts recovering to the equilibrium state. Here, the phase front separating the photo-disordered from the unperturbed region of the film begins shifting back towards the substrate interface, and at the same time the domain boundary further broadens.

The heterogeneous dynamics discussed above are unique to the mid-infrared excitation of the substrate lattice. Near-infrared illumination of the same heterostructure at 800 nm, which involves significant homogeneous charge excitation in the nickelate film [30], shows that in this case the magnetic order is uniformly melted over the entire film. This can be clearly deduced from Figs. 5(a) and (b) unveiling that the 800-nm excitation reduces the intensity of the magnetic (1/4 1/4 1/4) diffraction peak and of the Laue oscillations, without any noticeable change in the peak width. The same fitting procedure as applied above for the mid-infrared substrate excitation (see also the red solid lines in Figure 5(b)), shows the uniform and instantaneous decrease of the magnetic order parameter across the whole film thickness.

We next turn to a discussion of the possible physical mechanisms underlying these observations. Firstly, we can exclude direct absorption of the mid-IR pump pulse in the nickelate film as being responsible for the observed magnetization dynamics. This is not only because the energy deposited in the nickelate is negligible (3% of the incident energy), but also because the 1-µm penetration depth in $NdNiO_3$ would imply homogeneous excitation. Secondly, as the speed of the magnetic melt front is as high or higher than the nominal longitudinal sound velocity [33], heat slowly propagating from the substrate cannot be causing the effect observed here. As a third possibility, one could consider a structural phase transition triggered at the buried interface, propagating inward and driving the magnetic melt front. However, a structural front propagates far slower than the speed of sound and is therefore also unlikely [34]. Finally, a propagating acoustic wave is also unlikely, as it is not clear how the resonant excitation of a q~0 optical phonon could launch a propagating strain pulse on this short time scale, and the strain extracted from the shift in the diffraction peak is never larger than $10^{-3}$.

We propose that the direct excitation of the infrared-active substrate phonon mode induces octahedral distortions [35,36,37] across the interface to locally act on the electronic and magnetic ordering of the nickelate film. Indeed, similar scenarios have been predicted for perovskite heterostructures in the static case [38]. Propagation of the phase would then be driven by electronic rather than magnetic effects. We note that the anisotropic magnetic interaction stabilizing the Neel order introduce a spin gap and hence flatten the dispersion of magnons (see Supplemental Material for details), resulting in reduced group velocity and magnon localization.

We believe that charge carriers, which are made mobile at the interface and are very efficient at randomizing spin correlations [39], are more likely to drive the magnetic front. As these itinerant carriers acquire kinetic energies that exceed magnetic energy scales, these can sustain a front that propagates. The scrambling of antiferromagnetic order would transfer initial kinetic energy into the magnetic sector [40], eventually leading to a stalling of the phase front. Although a mechanistic description of this process requires materials specific calculations [41,42], which are beyond the scope of the present paper, a model Hamiltonian description, in which charges are freed at the interface and are mobile through diffusion, confirms these qualitative arguments (see Supplemental Material). This scenario is also compatible with the similar time scales observed for the insulator-metal transition and the magnetic order melting presented in Fig. 1(d).

In summary, we have shown that the dynamics of magnetic disordering in complex oxide heterostructures follows distinctly new physical pathways when the substrate lattice is excited. We demonstrate that magnetic melting is initiated at the buried interface and propagates into the film at speeds of order of or higher than the speed of sound. We assign the underlying physics to an electronic itinerancy front, which carries spin disordering and possibly local lattice distortions. The ability to control such ultrafast magnetic phase fronts with light may be conducive to new applications in optomagnetic devices, in which information may be encoded and shuttled in domain walls at faster rates than more established spin torque driven walls [43].

## ACKNOWLEDGEMENT

We thank A. Frano for helpful discussions.

Portions of this research were carried out on the SXR Instrument at the Linac Coherent Light Source (LCLS), a division of SLAC National Accelerator Laboratory and an Office of Science user facility operated by Stanford University for the U.S. Department of Energy. The SXR Instrument is funded by a consortium whose membership includes the LCLS, Stanford University through the Stanford Institute for Materials Energy Sciences (SIMES), Lawrence Berkeley National Laboratory (LBNL, contract No. DE-AC02-05CH11231), University of Hamburg through the BMBF priority program FSP 301, and the Center for Free Electron Laser Science (CFEL).

The research leading to these results has received funding from the European Research Council under the European Union's Seventh Framework Programme (FP7/2007-2013) / ERC Grant Agreement n° 319286 (Q-MAC) and n°281403 (FEMTOSPIN). Work performed at SIMES was further supported by U.S. Department of Energy, Office of Basic Energy Science, Division of Materials Science and Engineering, under the Contract No. DE-AC02-76SF00515. Work at Brookhaven National Laboratory was funded by the Department of Energy, Division of Materials Science and Engineering, under contract No. DE-AC02-98CH10886.

## AUTHOR CONTRIBUTIONS

A.D.C., M.F. and A.C. conceived this project. M.F., A.D.C., R.M., V.K., S.B.W., S.S.D., and J.P.H performed the experiment at the LCLS, supported by W.F.S., J.J.T. and G.L.D (beamline), M.P.M. and J.R. (laser), Y.D.C and W.S.L. (experimental endstation). The sample was grown by R.S., P.Z., and J.M.T.. M.F. analyzed the data with help from H.B..

# FIGURE CAPTIONS

**Fig. 1. Magnetic ordering in NdNiO$_3$ and the vibrationally induced phase transition**
(a) Visualization of the complex antiferromagnetic ordering in NdNiO$_3$. (b) Energy scan of the (1/4 1/4 1/4) diffraction peak, associated with the magnetic order on the Ni sublattice, around the Ni L$_{2,3}$ edges. These data are measured on the NdNiO$_3$/LaAlO$_3$ heterostructure. (c) Wavelength dependent penetration depth of mid-infrared light around the NdNiO3 (green) and LaAlO3 (blue) high-frequency phonon resonances. The energy spectrum of the excitation pulses at 15 μm wavelength (black data points, fitted by a Gaussian profile in red) is tuned in resonance with the substrate mode. (d) Blue: Changes of the peak intensity of the Ni L$_3$ diffraction peak at 852 eV, following this optical excitation with a fluence of 4 mJ/cm$^2$. The green data points show the transient THz reflectivity of the same sample under identical excitation conditions. The melting of the magnetic order and the insulator-metal transition both take place on the same time scale.

**Fig. 2: Temporal evolution of the antiferromagnetic order in reciprocal space** (a) Photo-induced dynamics of the momentum dependence of the (1/4 1/4 1/4) Ni L$_3$-edge diffraction peak (logarithmic scale), on both short and long time scales. (b) The same diffraction peak at selected time delays before and after the mid-infrared excitation, overlaid with numerical fits assuming a heterogeneous melting of the antiferromagnetic order triggered at the NdNiO$_3$/LaAlO$_3$ interface.

**Fig. 3: Real space dynamics of the antiferromagnetic order** (a) Schematic illustration of the demagnetization process. At negative time delay, the NdNiO$_3$ film is homogeneously ordered. Direct excitation of a substrate phonon triggers a magnetic melt front, propagating from the interface into the film. (b) Short time scale spatial dynamics

of the magnetic order parameter along the [111] direction, extracted from the numerical fits shown in Figure 2. (c) Centre position of the phase front, separating the melted from the unperturbed region of the film (blue dots). The striped area marks the sonic regime, given by the $4.1\times10^3$ m/s NdNiO$_3$ longitudinal speed of sound according to Ref. 33.

**Fig. 4: Recovery dynamics of the antiferromagnetic order** (a) Three-dimensional plot of the long time scale spatial dynamics of the magnetic order parameter. The recovery to the equilibrium state takes place on the time scale of 100 ps. (b) Time-dependent position and width of the magnetic phase front, extracted from the data shown in panel (a).

**Fig. 5: Magnetization dynamics following Ni charge transfer excitation at 800 nm** (a) Photo-induced dynamics of the (1/4 1/4 1/4) Ni L$_3$-edge diffraction peak in reciprocal space (logarithmic intensity scale). The excitation fluence is about 4 mJ/cm$^2$. (b) The same diffraction peak at selected time delays before and after the near-infrared excitation, together with numerical fits (red solid lines) of Eqs. (1) and (2). (b) Three-dimensional plot of the spatiotemporal dynamics of the magnetic order parameter, showing homogeneous photo-induced demagnetization.

**Figure 1**

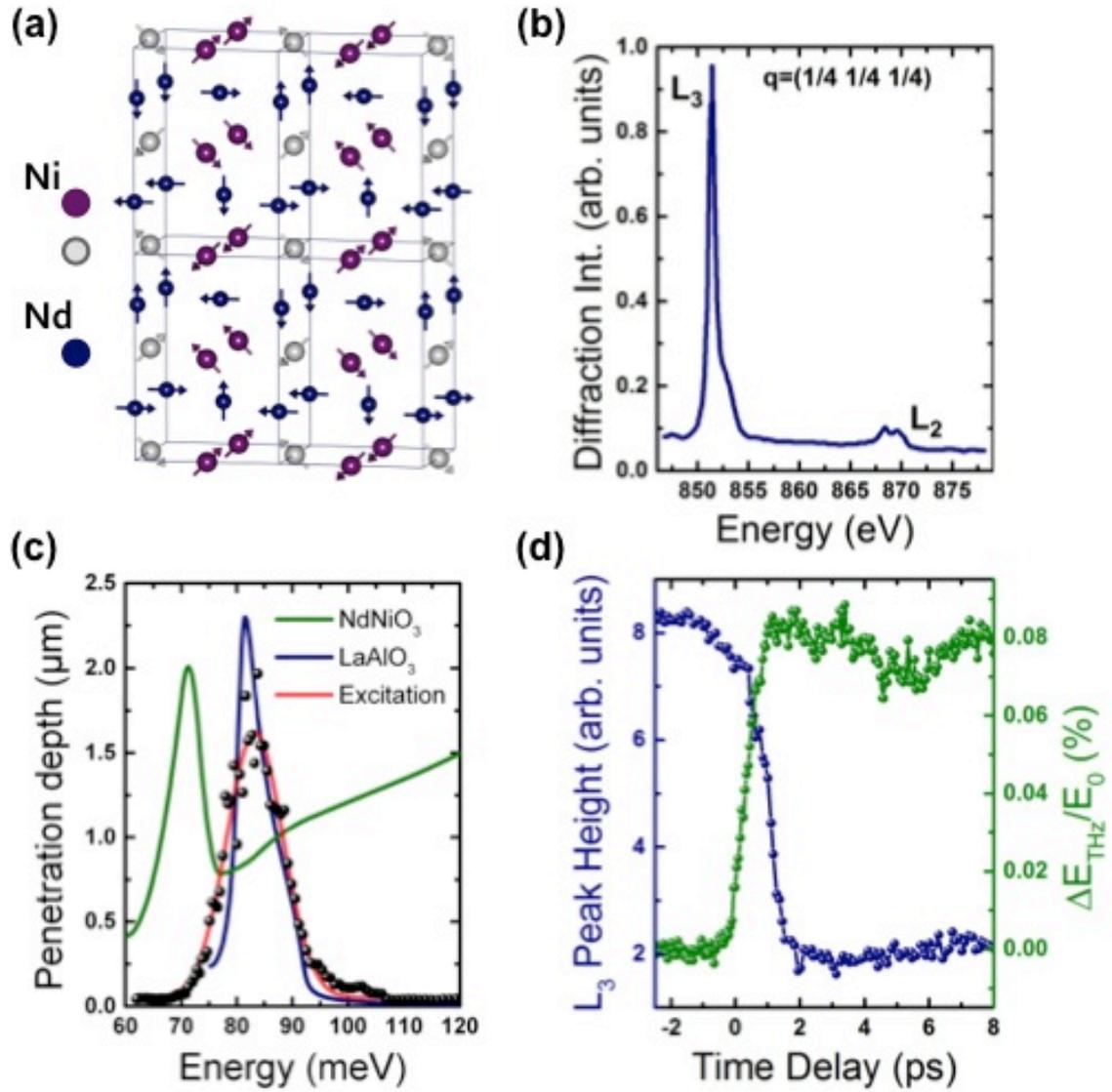

**Figure 2**

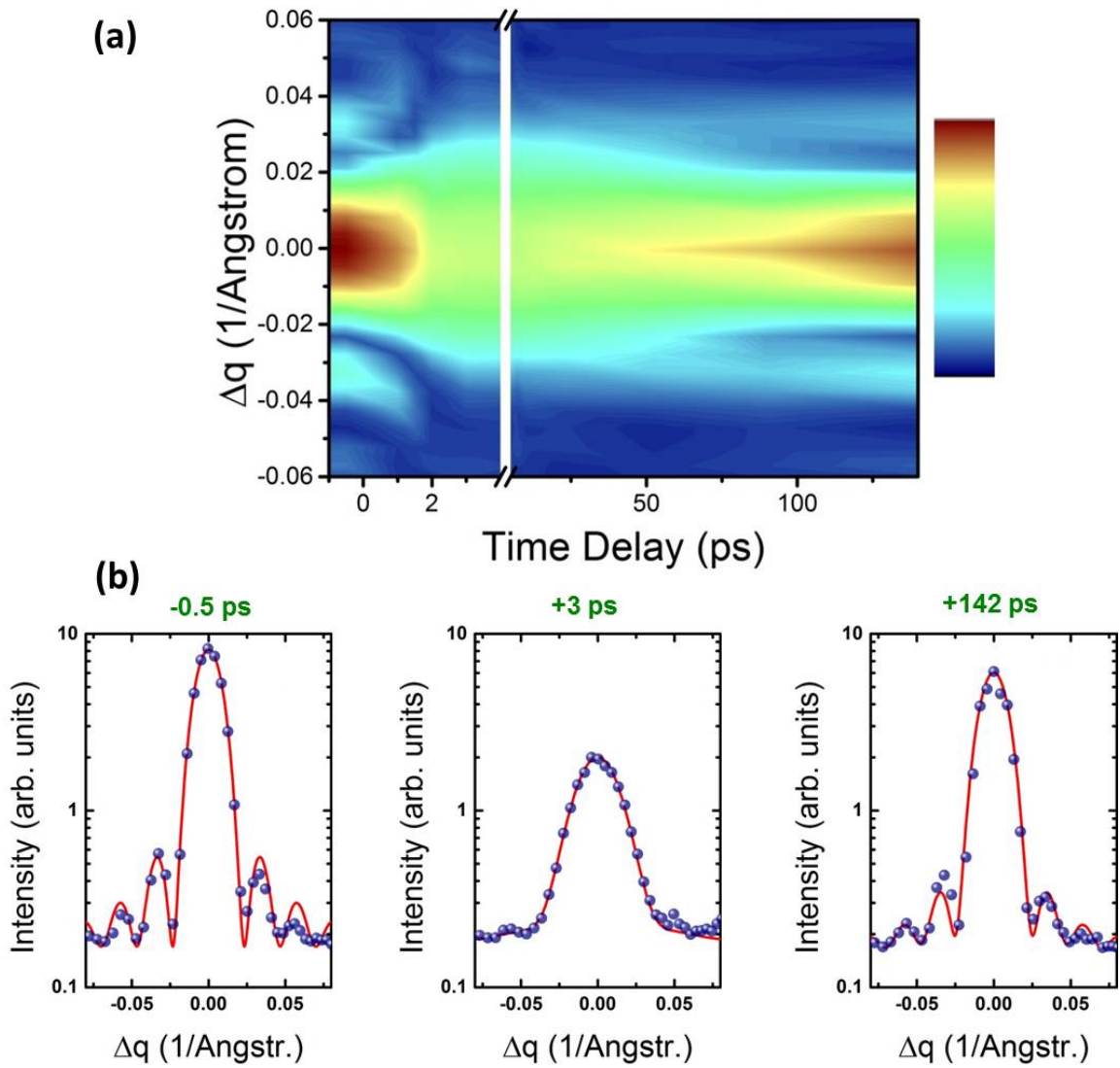

**Figure 3**

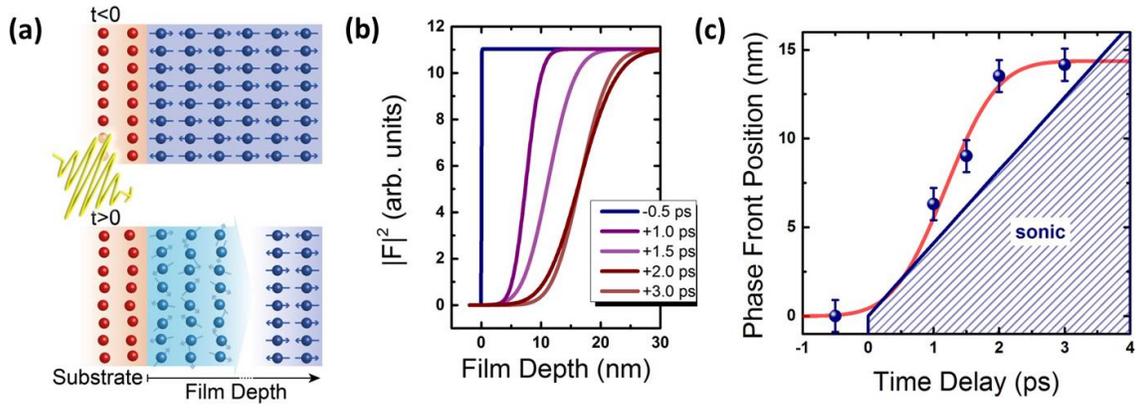

**Figure 4**

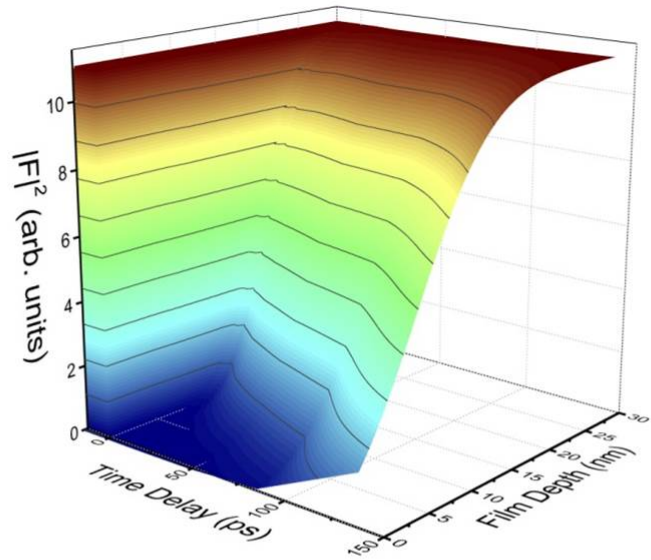

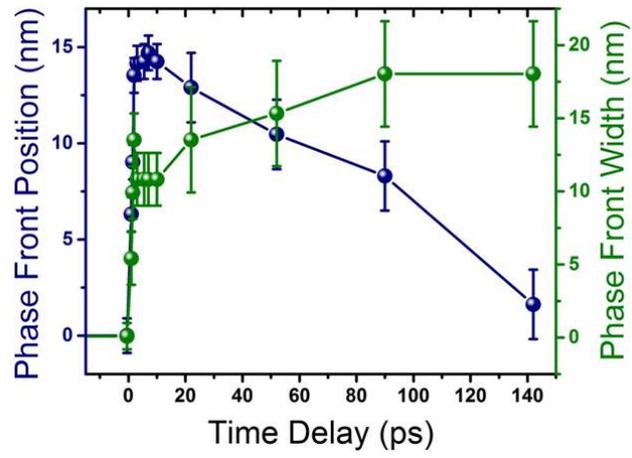

**Figure 5**

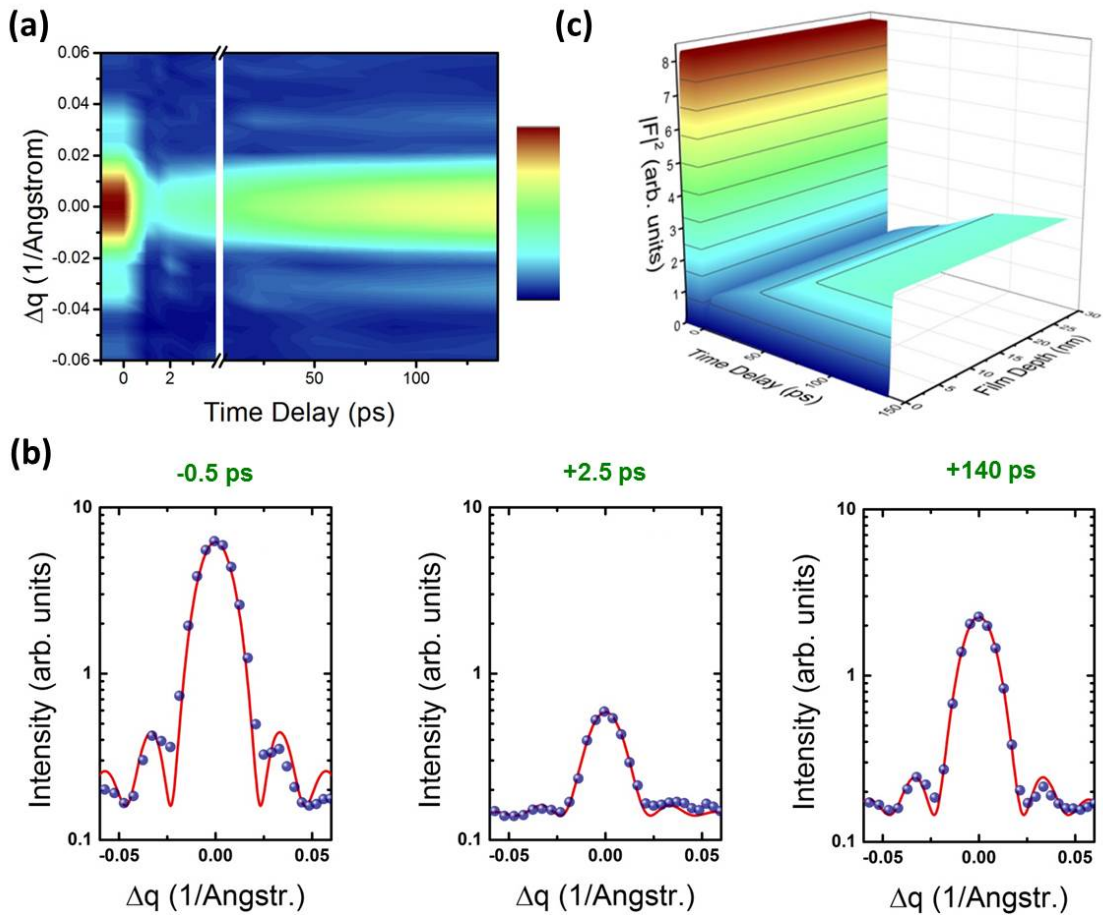

# Supplemental Material:
# Spatially resolved ultrafast magnetic dynamics launched at a complex-oxide hetero-interface

M. Först[1*], A.D. Caviglia[2*], R. Scherwitzl[3], R. Mankowsky[1], P. Zubko[3], V. Khanna[1,4,5], H. Bromberger[1], S.B. Wilkins[6], Y.-D. Chuang[7], W.S. Lee[8], W.F. Schlotter[9], J.J. Turner[9], G.L. Dakovski[9], M.P. Minitti[9], J. Robinson[9], S.R. Clark[10,4], D. Jaksch[4,10], J.-M. Triscone[3], J.P. Hill[6], S.S. Dhesi[5], and A. Cavalleri[1,4]

[1]*Max-Planck Institute for the Structure and Dynamics of Matter, Hamburg, Germany*

[2]*Kavli Institute of Nanoscience, Delft University of Technology, The Netherlands*

[3]*Département de Physique de la Matière Condensée, University of Geneva, Switzerland*

[4]*Department of Physics, Clarendon Laboratory, University of Oxford, UK*

[5]*Diamond Light Source, Chilton, Didcot, United Kingdom*

[6]*Condensed Matter Physics and Materials Science Department, Brookhaven National Laboratory, Upton, NY*

[7]*Advanced Light Source, Lawrence Berkeley Laboratory, Berkeley, CA*

[8] *The Stanford Institute for Materials and Energy Sciences (SIMES), Stanford Linear Accelerator Center (SLAC) National Accelerator Laboratory and Stanford University, Menlo Park, CA*

[9] *Linac Coherent Light Source, Stanford Linear Accelerator Center (SLAC) National Accelerator Laboratory, Menlo Park, CA*

[10] *Centre for Quantum Technologies, National University of Singapore, Singapore*

**S1. Boundary induced melting of magnetic order**

Capturing the connections between spin and charge dynamics in nickelates invariably requires a multi-band *d-p* orbital Hubbard-like model, in which full degeneracy of the Ni 3*d* orbitals and the oxygen 2*p* orbitals, as well as their hybridization, is taken into account [1,2]. While the precise mechanistic description of the vibrational excitation involves such material specific calculations, here as a first step, we explore the generic effects of boundary excitations in a simpler model Hamiltonian.

Model Hamiltonian

To qualitatively describe the dynamics of the nickelate thin film we consider a single-band Hubbard model-like Hamiltonian [3] of the form

$$\hat{H} = -t \sum_{\langle i,j \rangle} \left( \hat{c}_{i\sigma}^+ \hat{c}_{j\sigma} + h.c. \right) + U \sum_j \hat{n}_{j\uparrow} \hat{n}_{j\downarrow} + J_{zz} \sum_{\langle i,j \rangle} \hat{s}_i^z \hat{s}_j^z$$

where $\hat{c}_{i\sigma}^+$ is the creation operator for an electron of spin $\sigma$ on site $i$, $\hat{n}_{i\sigma}$ is the corresponding number operator and $\hat{s}_i^z = \frac{1}{2}(\hat{n}_{i\uparrow} - \hat{n}_{i\downarrow})$ is the spin projection along the $z$ axis (out-of-plane direction). We denote nearest-neighbor sites in the lattice as $\langle i,j \rangle$, $t$ as the hopping amplitude, $U$ as the on-site Coulomb repulsion, and $J_{zz}$ as an additional Ising spin interaction strength. At half-filling, and in the strong-coupling limit $U/t \gg 1$, the dynamics of the spin sector arising from $\hat{H}$ is an XXZ model

$$\hat{H}_{XXZ} = J \sum_{\langle i,j \rangle} \left( \hat{s}_i^x \hat{s}_j^x + \hat{s}_i^y \hat{s}_j^y + \Delta \hat{s}_i^z \hat{s}_j^z \right),$$

where $J = 4t^2/U$ and $\Delta = 1 + J_{zz}/J$. Since the nickelate film displays antiferromagnetic ordering with a well defined orientation we have included the Ising interaction $J_{zz} > 0$ which breaks the rotational invariance of the $\Delta = 1$ isotropic Heisenberg model. For $J_{zz}/J \gg 1$ the ground state of the XXZ model has classical Ising character akin to a simple Neel state $\uparrow \downarrow \uparrow \cdots \downarrow \uparrow$ modified by in-plane spin-flip fluctuations.

Within this model framework we now investigate how magnetic and electronic dynamics nucleated at the boundary of the film can propagate through the system, and the effects their motion has on the magnetic ordering.

Magnetic dynamics

One possibility is that an interfacial quench of the film generates localized spin-flips of the antiferromagnetic order, which subsequently evolve according to the XXZ

Hamiltonian $\hat{H}_{XXZ}$. After performing a Holstein-Primakoff transformation [4] and retaining only the lowest order terms an antiferromagnetic spin-wave (magnon) dispersion relation is obtained as [5]

$$\epsilon(\vec{q}) = \eta \frac{J}{2}\sqrt{\Delta^2 - \gamma(\vec{q})},$$

where $\vec{q}$ is the quasi-momentum in a bipartite lattice with coordination number $\eta$ and $\gamma(\vec{q}) = \frac{1}{\eta}\sum_{\vec{a}} e^{i\vec{q}\cdot\vec{a}}$ with $\vec{a}$ being the real space vectors connecting one site to all its nearest neighbors. In 3D simple cubic lattice with spacing $a$ we have that $\gamma(\vec{q}) = \frac{1}{3}[\cos(q_x a) + \cos(q_y a) + \cos(q_z a)]$.

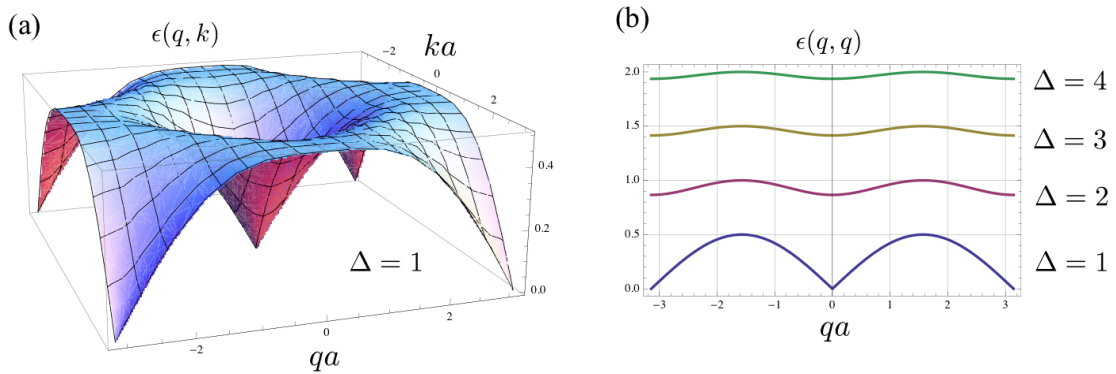

**Fig. S1:** (a) The antiferromagnetic spin-wave dispersion relation $\epsilon(q,k)$ for the XXZ spin model in a 2D square lattice at the isotropic Heisenberg limit $= 1$. (b) Diagonal cuts through of the dispersion relation $\epsilon(q,q)$ for different values of $\Delta \geq 1$.

In Fig. S1(a), the dispersion $\epsilon(q,k)$ for a 2D square lattice is shown for $\Delta = 1$ and presents a gapless mode $\vec{q} = (0,0)$ around which the spectrum is linear $\epsilon(\vec{q}) \sim |\vec{q}|$. An additional zero mode is present at the Brillouin zone edge $\vec{q} = (\pi, \pi)$, corresponding to the antiferromagnetic ordering wave-vector, as direct result of having broken the continuous rotational symmetry. However, the more relevant situation for the nickelate film is when an anisotropy $\Delta > 1$ is present. As can be seen in Fig. S1(b), an increasing $\Delta$ quickly introduces a gap at all quasi-momentum $\vec{q}$ and flattens $\epsilon(\vec{q})$ leading to a suppression of the spin-wave group velocity. This reflects the detuning of spin-flip transitions with amplitude $J < J_{zz}$ which control the motion.

As a result the propagation of localized magnon packet in real space, corresponding to coherent superposition of all spin-wave momenta, will be substantially reduced by the anisotropy. Thus, magnetic dynamics nucleated at the boundary will remain localized there. This is readily confirmed by time-dependent mean-field calculations. Further to this lack of propagation, for strong anisotropies, the XXZ model is also known to possess highly excited edge-locked bound states which pin ferromagnetic regions to open boundaries [6].

The purely magnetic process of melting the bulk antiferromagnetic order by boundary excitations is also not favored energetically. Magnons possess an energy governed by, and therefore the same order as, the interaction which stabilizes the order. Assuming that the total excitation energy of the boundary magnons scales with the interfacial area, this will never be energetically sufficient to cause the complete melting of the antiferromagnetic order in a fraction of the bulk.

Charge dynamics

We now consider the consequences of the substrate vibration nucleating charge-neutral holon (vacancy $|0\rangle$) and doublon (double occupancy $|\updownarrow\rangle$) pairs at the boundary of the film. Compared with a "magnetic only" picture, this scenario, where itinerant charge carriers are generated, is also consistent with the concomitant insulator-metal transition observed in the THz probe experiment.

To describe the dynamics of the system when doped with charge carriers requires the full Hubbard Hamiltonian $\widehat{H}$ which now involves two new energy scales $U$ and $t$ not previously exposed in $\widehat{H}_{XXZ}$. Once $U$ is sufficiently large then regime $t > J_{zz} >$

$4t^2/U$ is attained where the anisotropy still dominates the magnetic interactions while hopping exceeds all magnetic energy scales. Since the kinetic energy of localized holon-doublon pairs is $\sim t$ this indicates that, in contrast to magnons, charge carriers can be energetic enough to both propagate and significantly excite the magnetic sector.

The motion of holons or doublons in an antiferromagnet can scramble the antiferromagnetic order creating a paramagnet. At $T = 0$, this is a long-studied problem [7,8,9,10]. Key insight is that the motion of charge carriers is greatly impeded by the presence of antiferromagnetic order. Approximations, such as Brinkman and Rice's retraceable path approach [7], indicate the motion of charge carriers is akin to Brownian motion even at $T = 0$. Moreover, due to the 'string' magnetic excitations, which trail behind the trajectory of a charge in an antiferromagnet, it is expected that their propagation will be confined to a finite region to account for the loss of kinetic energy to the magnetic sector [9,10].

We introduce an approximate classical stochastic model to capture these effects based on a Pauli master equation [11]

$$\frac{d}{d\tau}P_j(\tau) = \sum_{k \neq j} W(k \to j) P_k(\tau) - \sum_{k \neq j} W(j \to k) P_j(\tau),$$

describing the evolution in time $\tau$ of probability $P_j(\tau)$ of occupying a particular configuration $j$ of spins are charges in the Hubbard lattice, e.g. $|j\rangle = |\uparrow\downarrow\updownarrow\downarrow\uparrow\downarrow\,0\,\uparrow\downarrow\cdots\rangle$. The transition rate $W(k \to j)$ from configuration $k$ to $j$ is then defined as

$$W(k \to j) = \begin{cases} w, & E_j - E_{\text{init}} \leq \mathcal{E} \\ 0, & \text{otherwise} \end{cases},$$

where $w$ is a constant rate, $E_j$ is the energy of configuration $j$ and $E_{\text{init}}$ is the energy for the initial boundary excited configuration evaluated with the Hubbard model when hopping $t = 0$, and $\mathcal{E}$ is the total excitation energy. Carriers therefore diffuse through the lattice at a fixed rate $w$ until the energy deposited into system by moving to a new configuration exceeds $\mathcal{E}$, at which point the rate abruptly drops to zero. This simple choice enforces a form of energy conservation which limits the capacity of the holons and doublons to excite the magnetic sector.

The resulting stochastic evolution starting from an antiferromagnetic configuration with holon-doublon pairs randomly distributed along the boundary was simulated using dynamical Monte Carlo [12]. A 40×40 2D square lattice was used with periodic boundaries along the interface $x$ direction and open boundaries in the film $z$ direction. In Fig. S2(a), we report the staggered magnetization averaged over the $x$ direction as a function of the $d_z$ distance into the film and time. Specific time slices are also displayed in Fig. S2(b). Together, these show that initially there is fast evolution of an error-function diffusion front dictated by the rate $w$ which strongly demagnetizes the lattice behind it. The specific parameters $\mathcal{E}$ and $w$ used (see figure caption) were chosen to best match the features seen in the experimental data. Consequently after ~3 ps the diffusion front abruptly stalls, due to the energy limit $\mathcal{E}$, corresponding to when antiferromagnetic order has been melted in approximately half the film. This emergence of a gradual boundary between paramagnetic and antiferromagnetic regions in the film is further highlighted by the dynamical suppression of the $qa = \pi$ peak in the static structure factor

$$F(q) = \sum_{jl} e^{iq(j-l)a} \langle \hat{s}_j^z \hat{s}_l^z \rangle,$$

shown in Fig. S2(c) as a function of time.

While this model has ad-hoc features it nonetheless illustrates that diffusive motion of charge carriers is a highly effective at scrambling antiferromagnetic order simply by the shuffling of spins along a trajectory as shown in Fig. S2(d). The further inclusion of an energy cut-off $\mathcal{E}$ however was essential to capture the localization effects [9,10] caused by the loss of kinetic energy of the charge carriers as they excite the magnetic degrees of freedom, also shown in Fig. S2(d). Both these effects appear to be qualitatively present in the experimental results.

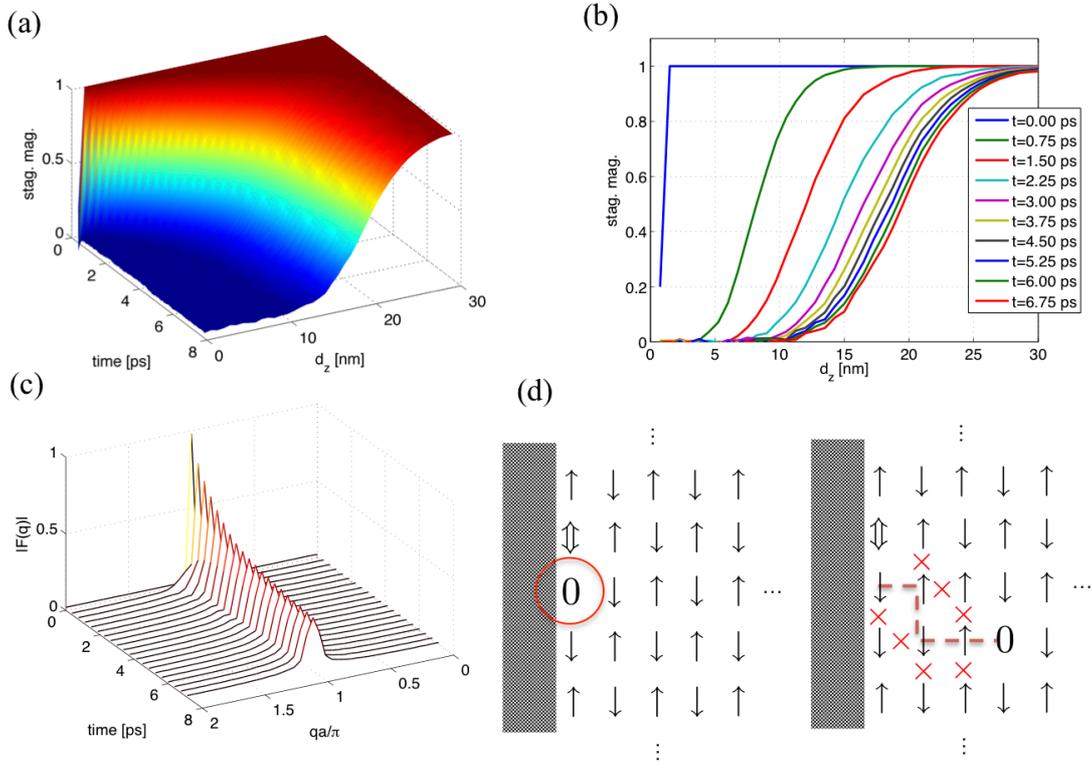

**Fig. S2** (a) The staggered magnetization for a 2D square lattice as a function of distance from the interface (film depth) and time averaged over the transverse direction. (b) Slices of plot (a) for given times. (c) The magnetic structure factor $F(q)$, for the lattice as function of time. To produce results quantitatively resembling the experimental data these calculations used a total of 32 holon-doublon pairs initiated with an excitation energy $\mathcal{E}$ equivalent to $\sim 47 J_{zz}$ per charge carrier and a diffusion time constant $w^{-1} \sim 38$ fs. (d) An illustration of how the stochastic motion of a charge carrier from the boundary (hole in this case) scrambles the antiferromagnetic order of the system. The dotted line denotes its trajectory and the ×'s mark the magnetic interactions which are now ferromagnetic bonds.

## S2. In-plane magnetization dynamics following mid-IR excitation

Here, we present the mid-IR induced dynamics of the antiferromagnetic in-plane correlations. Figure S3 shows the transverse rocking curves (theta scans) measured at three early time delays across the phase transition. The experimental conditions are identical to those presented in the main paper. Fitting Gaussian functions to these data sets yields FWHM values of 0.65±0.1 degrees (at –0.5 ps), 0.66±0.1 degrees (at +1.5 ps) and 0.63±0.1 degrees (at +3 ps). Thus, we conclude the in-plane correlation length remains unchanged. This effect results from the pump spot size being far larger than the magnetic domain size along this direction.

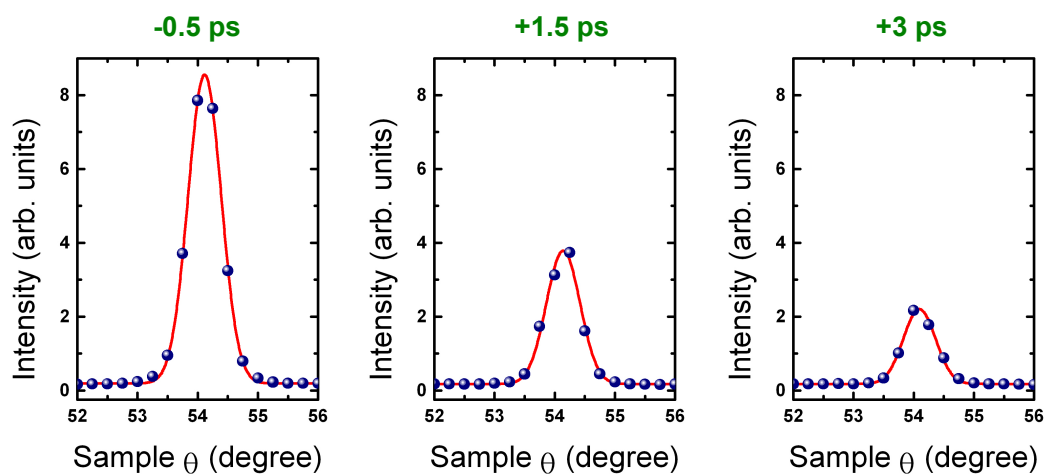

**Fig. S3** Rocking curves of the (1/4 1/4 1/4) diffraction peak at selected, early time delays before and after the mid-infrared excitation. The widths of these peaks represent the in-plane correlation length. Note, at 3 ps time delay, the system has undergone the demagnetization process along the out-of-plane direction, before it starts recovering.